\documentclass[aps, reprint, superscriptaddress, nofootinbib]{revtex4-1}
\usepackage{amsmath, latexsym, amssymb, hyperref, graphicx, color, slashed}
\definecolor{nicered}{rgb}{0.7,0.1,0.1}
\definecolor{nicegreen}{rgb}{0.1,0.5,0.1}
\hypersetup{colorlinks, citecolor=nicegreen,linkcolor=nicered}

\begin{document}

\newcommand{\Sec}[1]{ \medskip \noindent {\sl \bfseries #1}}
\newcommand{\subsec}[1]{ \medskip \noindent {\sl \bfseries #1}}
\newcommand{\Par}[1]{ \medskip \noindent {\em #1}}

\title{ Probing Seesaw  with  Parity Restoration }

\author{Goran Senjanovi\'{c}}
\affiliation{Gran Sasso Science Institute, Viale Crispi 7, L'Aquila 67100, Italy}
\affiliation{International Centre for Theoretical Physics, Trieste 34100, Italy }
\author{Vladimir Tello}
\affiliation{Gran Sasso Science Institute, Viale Crispi 7, L'Aquila 67100, Italy}
\date{\today}
\date{\today}

\begin{abstract}

 We present a novel way of testing the seesaw origin of neutrino mass in the context of the minimal Left-Right Symmetric Model. It is based on the connection between the leptonic interactions of the doubly charged scalars, whose presence is at the core of the seesaw mechanism, and the neutrino Dirac Yukawa couplings which govern, among other processes, the right-handed neutrino decays into left-handed charged leptons. We prove that any physical quantity depending on these couplings is a function of the hermitian part only which can significantly simplify their future experimental determination.

   \end{abstract}

\maketitle

\Sec{I. Introduction.} 
 The Standard electroweak Model (SM) started as a gauge theory of weak interactions and over the years, with the advent of the Higgs mechanism, turned  into a theory of particle masses.  In the case of charged fermions this mechanism can be verified from the correlation between their masses  and the Higgs boson decays into fermion antifermion pairs. 

   A key task of present-day particle physics is to achieve similar correlations for neutrino.
    A natural candidate  to address this issue   is the minimal left-right (LR) symmetric theory~\cite{PatiSalam}, suggested originally to explain parity violation in weak interactions through the spontaneous breaking of LR symmetry, which led to a non-vanishing neutrino mass and eventually to the seesaw mechanism~\cite{Minkowski, Mohapatra:1979ia, rest}  as an explanation of its smallness.  The  Majorana nature of heavy right-handed (RH) neutrinos leads to  the production of same-sign lepton pairs at hadronic colliders~\cite{Keung:1983uu} 
     and to a new contribution~\cite{Mohapatra:1979ia,MohSenj81} to neutrinoless double beta decay, two deeply interconnected manifestations of lepton number violation \cite{Tello:2010am}.
      This framework provides an ideal setting to address the issue of  neutrino mass.

Indeed, 
we find the existence  of a novel connection
between the neutrino Dirac Yukawa couplings and  the leptonic doubly charged scalar interactions which allows to predict a number of physical processes.
This way of disentangling the seesaw mechanism
 emerges as a consequence of generalized parity ($\mathcal{P}$) as the LR symmetry  and goes along the same lines as the correlation between the left-handed (LH) and RH quark mixings \cite{Senjanovic:2014pva}.

   For the case of generalized charge conjugation as the LR symmetry the above connection is lost. In this case, however,  it is possible to determine the neutrino Dirac Yukawa couplings from the knowledge of the LH and RH neutrino masses and mixings~\cite{Nemevsek:2012iq}. Still, in the case of  $\mathcal{P}$ chosen here we show that such a way of unravelling the seesaw mechanism could also be achieved. We demonstrate this explicitly for purely hermitian Dirac Yukawa couplings, although the employed procedure  can serve to tackle the general case as well.  Needless to say, the central aspect of our work, i.e., the correlation between the Dirac mass matrix and the doubly charged scalar Yukawa couplings, is valid in all of the parameter space. 
   \\

\Sec{II. The Minimal LR Model.}
The minimal left-right symmetric model (LRSM) is based on the gauge group $SU(2)_L \otimes SU(2)_R \otimes U(1)_{B-L}$, enlarged by generalized parity $\mathcal{P}$. Among other features, this implies the equality of gauge couplings $g_L = g_R \equiv g$. Fermions  transform as LR symmetric doublet representations $q_{L,R} = (u,d)_{L,R}$, 
$\ell_{L} = (\nu_L, e_L)$,  $\ell_{R} = (N_R, e_R)$ and the charged gauge interactions are given by
\begin{align} \label{eqWLWR}
 & \mathcal L_{gauge} =  \frac{g}{\sqrt 2} \left( 
  \overline \nu_L V_{L}^\dag \slashed{W}_{\!L} e_L + 
  \overline N_R  V_{R}^\dag \slashed{W}_{\!R} e_R\right) + \text{h.c.} 
\end{align}
where $V_L$ is the PMNS leptonic mixing matrix and $V_R$ its RH analog.

 Once produced, the RH charged gauge boson $W_R$ decays into RH neutrinos which allows to determine their masses and mixings through the so-called KS process~\cite{Keung:1983uu}. This is tailor made for hadronic colliders, such as the LHC where the experimental reach goes all the way up to a $W_R$ mass of about 6 TeV~\cite{Ferrari:2000sp}.

The Higgs sector consists~\cite{Minkowski,Mohapatra:1979ia,MohSenj81} of a complex bi-doublet $\Phi(2,2,0)$ and complex triplets 
$\Delta_L(3,1,2)$ and $\Delta_R(1,3,2)$ with quantum numbers referring to $SU(2)_L \otimes SU(2)_R \otimes U(1)_{B-L}$. 

At the first stage of symmetry breaking, the neutral component of $\Delta_R$ develops a vev $v_R$ and breaks the original symmetry down to the SM one. The latter is broken through the vevs of the bi-doublet neutral components 
\begin{equation}
\langle\Phi\rangle=v\, \text{diag} (\cos\beta,-\sin\beta e^{-ia}).
\end{equation}
%
The small parameter $s_at_{2 \beta}$ measures the amount of spontaneous CP violation and is bounded from 
 above~\cite{Senjanovic:2014pva}
 by
 \begin{equation} \label{epsilonlimit}
    s_at_{2 \beta} \lesssim  2 m_b/m_t
    \end{equation}
which makes it a suitable expansion parameter.

The electroweak symmetry breaking turns on a small vev for $\Delta_L$, $v_L \propto v^2/v_R$~\cite{MohSenj81}, whose experimental determination is somewhat involved~\cite{Perez:2008ha}.

   The  lepton Yukawa couplings in the minimal theory take the following form 
\begin{align}\label{eq:quarks&phi}
- &\mathcal L_Y=  \overline{\ell_{L}}\,\big(Y_{1} \Phi - Y_{2}\,  \sigma_2 \Phi^* \sigma_2)\, \ell_R \nonumber
\\ &
+\frac{1}{2} \ell^T_L Y_{L}  i\sigma_2 \Delta_L \ell_L +\frac{1}{2} \ell^T_R Y_R   i\sigma_2 \Delta_R\ell_R + \text{h.c.}
\end{align}

Under generalized parity  the fields transform as $\ell_L \leftrightarrow \ell_R$, $\Phi \to \Phi^{\dagger}$, $\Delta_L \leftrightarrow \Delta_R$, which implies  the following relations between Yukawa matrices 
\begin{align}\label{eq:yuk-hermitian}
Y_{1,2}=Y_{1,2}^{\dagger},\qquad Y_{L}=Y_{R}.
\end{align}
These equalities play an essential basis in what follows.

\Sec{III. Lepton masses and mixings.} 
 From \eqref{eq:quarks&phi}, the hermitian nature of $Y_{1,2}$ implies the following relations 
\begin{align}
\label{eq1}
 &M_D- U_{e}M_D^{\dagger}U_{e}=is_at_{2\beta}(e^{ia}t_{\beta}M_D+m_{e})
 \\[8pt]
 \label{eq2}
  &m_{e}-U_{e} m_{e}U_{e}=-is_at_{2\beta}(M_D+e^{-ia}t_{\beta}m_{e})
 \end{align}
where $M_D$ is the neutrino Dirac mass matrix. The unitary matrix $U_e$ is given by $U_{e}=E_R^{\dagger}E_L$, where $E_{L,R}$ are the left and right-handed unitary rotations of charged leptons obtained from the diagonalization of their mass matrix $M_{e}=E_Lm_{e}E_R^{\dagger}$.

In the same basis the light neutrino mass matrix $M_{\nu}$ in the seesaw picture~\cite{MohSenj81} becomes
  \begin{align}\label{seesawP}
&M_{\nu}= \frac{v_L}{v_R}U_{e}^T  M_N^* U_{e}  - M_D^T    \frac{1}{M_{N}}   M_D,     
\end{align}
where $M_N$ is the mass matrix of heavy RH neutrinos. In terms of masses and mixings one has
\begin{align}\label{massmatrices}
M_{\nu}=V_L^*m_{\nu}V_L^{\dagger}, \,\,\,\,\,\,\, M_N=V_R m_N V_R^T
\end{align}

The presence of $U_e$ signals the lack of hermiticity of the charged lepton matrix due to the spontaneous breaking of parity. 
  To appreciate better what is going on, it is instructive to take the limit of the so-called type II 
  seesaw~\cite{typeII}
   in which the first term in the rhs  of \eqref{seesawP} dominates the neutrino mass. This does not necessarily imply a tiny $M_D$; it could be simply a consequence of  a large enough $v_L$. It is easy to show that, in this case, the masses of light and heavy neutrinos are proportional to each other $m_N \propto m_\nu$ and  $U_e = V_R V_L^{\dagger}$ up to an overall phase. Since in general the charged lepton matrix is not hermitian, the LH and RH lepton mixings are not necessarily equal, the difference of which is contained in the $U_e$ matrix.  
 
 In order to probe  the origin of neutrino mass in a generic seesaw, it would seem that one must determine $M_D$ from 
\eqref{seesawP},  hard to achieve since in general this matrix is not hermitian~\cite{NSTprogram}. 
   
    We offer here an alternative approach based on the following simple but important observation.
It turns out, from  \eqref{eq2}, that $U_e$ is not arbitrary but is actually correlated with the neutrino Dirac mass matrix 
\begin{align}\label{Ulformula}
U_{e}&=\frac{1}{m_{e}}\sqrt{m_{e}^2+is_at_{2\beta} (t_{\beta}e^{-ia}m_{e}^2+m_{e}M_D)} 
\end{align}
which  allows to disentangle the seesaw mechanism and leads to a number of important phenomenological implications.  

Since $U_e$ is unitary, $M_D$ is not an arbitrary complex matrix. Instead of $2 n^2$ it has only $n^2$ real elements. Indeed, it can be shown from \eqref{eq1} that its anti-hermitian part $M_D^A$ is not independent but can be determined from the knowledge of the hermitian part $M_D^H$. At leading order in $s_at_{2\beta}$ it takes the  form
\begin{align}\label{antihermitianMD}
M_D^{A}\!=\!\frac{is_at_{2\beta}}{2}\!\Big(\! m_{e}\!+\!2t_{\beta}M_D^H\!+\!\mathcal{H_D}M_D^H\!+\!M_D^H\mathcal{H_D} \!\Big) \!
\end{align}
where we have introduced the hermitian matrix
\begin{equation}
(\mathcal{H}_D)_{ij}= \frac{\,\,\,(M_D^H)_{ij}}{m_{e_i}+m_{e_j}}.
\end{equation}
   The fact that generalized parity  fixes $M_D^A$ simplifies matters tremendously and opens the possibility for the experimental determination of the full $M_D$ matrix, as we show below.

    A discussion is in order. First of all, in this work we  focus on a directly testable seesaw picture in which the RH neutrinos can actually be produced at the LHC or  next generation of hadronic colliders. This requires their masses to lie below 10 TeV or so.  From the seesaw formula \eqref{seesawP}, this means that the elements of $M_D$ are at most of the order  of MeV. This justifies the validity of the expansion in $s_at_{2\beta}$ and allows us to work at  leading order; for larger values of $M_D$  
   this would not be feasible a priori. 
   
   Very large $M_D$ becomes of special interest in grand unified theories where one expects superheavy RH neutrinos. This is beyond the scope of this Letter and  will be addressed  in a separate publication.
   
     We wish to emphasize that our formalism is general and can be used for any value of $M_D$. In principle, one could also imagine cancellations between  type I and type II 
  seesaw contributions, so that $M_D$ could be large even for relatively light 
     $N$. We do not advocate this, but  
     it can be dealt with in the same way without modification. 
 What matters is that expression  \eqref{Ulformula} for $U_e$ is valid for all $M_D$ satisfying \eqref{antihermitianMD}. On the other hand, in the absence of such cancellations,  $s_at_{2 \beta}$  is limited  by (3) just as in the quark sector.

    Next, in the same  verifiable seesaw picture, the RH neutrinos  born out of the production of $W_R$  lead to  a background free final state composed of charged lepton and jet pairs~\cite{Keung:1983uu}.
    They decay dominantly through a virtual $W_R$ which  allows to determine both
     $m_N$ and $V_R$~\cite{Das:2012ii}. 
     This process offers a starting point to probe the nature and origin of neutrino mass. 
   The subdominant decay of RH neutrinos 
     into LH leptons and RH antileptons allows also for the determination of $M_D$ by a careful measurement of the chirality of the outgoing charged lepton~\cite{Ferrari:2000sp, Han:2012vk}.
        Moreover, there are other ways of probing $M_D$ through, for example, the leptonic decay of the heavy scalar doublet $H$ belonging to the bi-doublet representation. Limits from $K$- and $B$-meson physics set a stringent limit on its mass of about $20$~TeV~\cite{OtherLR}.
          Once again,  only  the hermitian part $M_D^H$ is independent leading to a one-to-one correspondence between the number of independent channels of each of these decays and the real elements of the hermitian part.
   
  Finally, as we now show, with the experimental knowledge of  $M_D^H$  one can obtain the flavor structure of the LH doubly charged scalars and in this manner pave the way towards a verification of the Higgs origin of neutrino mass. 
   This is central to our work.

\Sec{IV. From Dirac to Doubly Charged.}
   The crucial point  to realize is that 
   the $U_e$ matrix  enters in the Yukawa interaction of the doubly charged scalars  $\delta_{L,R}^{++}$
            \begin{equation}
-\mathcal L_{\delta}\!=\!\frac{1}{2}   \, \delta_L^{^{++}}  e_{L}^T   \bigg(\!U_{e}^T \frac{M_N^*}{v_R} U_e\!\bigg)  e_L\!+ \frac{1}{2}\delta_R^{^{++}}  e_{R}^T \bigg(\!  \frac{ M_N^*}{v_R}\! \bigg)  e_R \!  
    \end{equation}

Since both LH and RH triplet Yukawa couplings  
  are proportional to the mass matrix of RH neutrinos,
the role of  $U_e$  is to account for the mismatch between LH and RH sectors.  

 Doubly charged scalars are produced pairwise and thus expected, unless very light,  to be less accessible than $W_R$. 
 The $\delta_R ^{++} \to e_{iR}^+ e_{jR}^+$ decays 
 provide a way of measuring $m_N$ and $V_R$, complementary to the KS process.  
 Even more interesting are the decays of $\delta_L^{++}\to e_{iL}^+ e_{jL}^+$  since they offer a  physical connection with the seesaw mechanism. 
  In addition to the RH neutrino masses and mixing,  their determination requires the knowledge of the $U_{e}$ matrix which probes $M_D$ through \eqref{Ulformula}.
By studying the flavor structure of  these decays and the corresponding low energy rare leptonic processes one can in principle extract  the elements of
the $U_e$ matrix.  We leave aside the low energy aspects and concentrate on high energy decays because colliders have a higher capacity to identify the chirality of outgoing leptons.

  In particular, the comparison between  $\delta_L^{++}$ and $\delta_R ^{++}$ decays is rather useful due to the common $m_N$ and $V_R$ dependence. After some deliberation one obtains, at leading order in small $ s_at_{2\beta}$, the following prediction for the  ratio between $\delta_L^{++}$ and $\delta_R^{++}$ decay rates into charged lepton pairs 
   \begin{equation}\label{doublychargeddecay}
  \frac{\Gamma_{\delta_L\rightarrow e_{L_i} e_{L_j}}}{\Gamma_{\delta_{R}\rightarrow e_{R_i} e_{R_j}}}
  \simeq \frac{m_{\delta_L } }{m_{\delta_R } } \!\bigg[1 
 \!+\!2 s_at_{2\beta} \, \text{Im} \frac{     \big( \mathcal{H}_D  M_N   +    M_N \mathcal{H}_D^T  \big)_{ij}    }{
  (M_N)_{ij}   }    \bigg]  
\end{equation}
%
As we argued before, the result manifestly depends only on the hermitian part of $M_D$.

Moreover, both $U_{e}$ and $M_D$ contribute to the neutrino mass. The correlation between these two quantities, previously thought unrelated, provides a novel way of disengaging the seesaw mechanism and  opens a new chapter into the probe of the origin of neutrino masses.

Besides the doubly charged states, the triplets contain also singly charged and neutral scalars. 
Their interactions are however of secondary importance since neutrinos are missing energy at colliders and we do not consider them here.

\subsec{ The hermitian case.} 
It is evident from \eqref{Ulformula} that $U_{e}$ becomes the identity matrix in the limit $s_at_{2\beta} \rightarrow 0$ and 
its connection with $M_D$ disappears. In this limit however one can express $M_D$ as a function of light and heavy neutrino masses and mixings.
 The point is that $M_D$ becomes purely hermitian and by using \eqref{seesawP}  it can be decomposed as 
 \begin{align}\label{MD-hermitian}
 M_D= V_R \sqrt{m_N}\, \left(O \sqrt{s}EO^{\dagger}\right) \,\sqrt{m_N} V_R^{\dagger}  
 \end{align}
 The orthogonal matrix $O$  and the symmetric normal form $s$ (see \cite{Gantmacher} for mathematical details) are obtained from the following symmetric matrix
  \begin{align}\label{Omatrix}
 \frac{v_L^*}{v_R}-\frac{1}{\sqrt{m_N}} V_R ^{\dagger }  V_Lm_{\nu}V_L^{T}  V_R^{*}\frac{1}{\sqrt{m_N}}= O s O^T
  \end{align}
 provided that the following conditions are satisfied
   \begin{align}\label{conditions}
\text{Im}\text{Tr} \left[ \frac{v_L^*}{v_R}- \frac{1}{M_N}  M_{\nu}^* \right]^n= 0,\quad (n=1,2,3)
  \end{align}
which imply that the phases of light and heavy neutrino mass matrices are not independent. 

It can be shown that due to the above constraints, excluding pathological situations, the symmetric normal form $s$ takes only two possible diagonal forms
\begin{equation}
  s_{I}=\text{diag}(s_1,s_0,s_2) ,\quad   s_{II}=\text{diag}(s,s_0,s^*) 
    \end{equation}
where $s_{0,1,2}$ are real, whereas $s$ is complex. 
 This in turn fixes uniquely the form of the matrix $E$ which ensures that  \eqref{MD-hermitian} is hermitian.  After some thought one gets\footnote{It is straightforward to obtain the matrix $E$ for the more complicated degenerate cases in which $s$ is not diagonal.} 
   \begin{equation}
E_{\text{I}}=  \left(   \begin{array}{ccc}
 1& 0&0 \\ 
  0& 1 & 0 \\ 
0 & 0 & 1
\end{array}   \right), \qquad  E_{II}=  \left(   \begin{array}{ccc}
 0& 0&1 \\ 
  0& 1 & 0 \\ 
1 & 0 &  0
\end{array}   \right)
    \end{equation}
 corresponding to $s_I$ and $s_{II}$, respectively. For 
 $s_a t_{2\beta}=0$, just as  in the case of charge conjugation~\cite{Nemevsek:2012iq}, 
  the Dirac Yukawa matrix
 depends only on the left- and right-handed neutrino mass matrices,  and takes a natural value on the order of $\sqrt{m_{\nu}m_N}$. 

 For the sake of illustration, we exemplify the above procedure for a simplified choice $V_R=V_L$ which 
 from \eqref{conditions} implies  a real $v_L$.  Using \eqref{MD-hermitian} the Dirac mass matrix then takes the simple form
  \begin{align}
 M_D=   V_L m_N \sqrt{ \frac{v_L}{v_R}-\frac{m_{\nu}}{m_N}}V_L^{\dagger} 
  \end{align}
  In general, for a non-vanishing $s_a t_{2\beta}$, there is an additional dependence on the masses of charged leptons as can be seen from \eqref{antihermitianMD}.

\subsec{V. Phenomenological implications.}
We discuss here how to probe  the origin of neutrino mass by determining the neutrino Dirac mass matrix and the doubly charged scalar decay rates. 
 
   As we already said, the Dirac Yukawa couplings can be obtain from rare  heavy $N$ decays into LH charged leptons or RH antileptons. These decays are due to the light-heavy neutrino mixing induced by the Dirac Yukawa couplings~\cite{Buchmuller:1991tu}.  The decay rate has the following form
   \begin{equation}
   \begin{split}
&\Gamma (N_i \to W_L^+ e^{}_{L_{ j}})\propto  \frac{  m_{N_i}}{M_{W_L}^2}
  \big|(V_R^{\dagger} M_D)_{ij}\big|^2
\end{split}
 \end{equation}
 Due to \eqref{antihermitianMD} this depends only on $M_D^H$.  In order for this channel to provide an efficient way of measuring the neutrino Dirac Yukawas, it needs the  $m_N$ and $V_R$  input from the KS process. 
In case $N$ were lighter than $W_L$, one would obviously have the analog $W_L \to N e_L$ decays controlled by the same couplings.

 With enough energy,   future colliders can also in principle detect the heavy scalar doublet $H$ and measure its leptonic decays.  In particular, the decays of the neutral component $H^0$ into charged lepton pairs are given by 
\begin{align}\begin{split}
\Gamma(H^0\rightarrow e_i\bar{e}_j)
\propto \frac{m_H}{M_{W_L}^2}  |(M_D^H+s_{2\beta}m_{e})_{ij}|^2
\end{split}
\end{align}
where for the sake of simplicity and illustration we have neglected the phase $a$. We have also ignored the tiny scalar mixings, suppressed by the electro-weak scale, between the superheavy $H^0$ and other scalars such as the SM Higgs and $\delta_R^0$. 
In this approximation, the real and imaginary components of $H^0$ can be treated as degenerate particles. 

These decays provide a rather clean test of $M_D^H$ since there is no additional lepton mixing. 
  Moreover, the flavor changing decays are directly proportional  to the off-diagonal elements of $M_D^H$ which facilitate their determination. It is important that there are nine different channels for nine different $M_D^H$ elements. 
  Of course, together with the above $N$ decays, one will end up with an over constrained system and be able to determine also $a$ and $\beta$.

The $a$ and $\beta$ parameters parameters enter in a number of other physical processes. For example, the mixing  $\xi_{LR}$ between $W_L$ and $W_R$ gauge bosons is given by $|\xi_{LR}| \simeq {M_{W_L}^2}/{M_{W_R}^2} \sin 2\beta$, which in principle allows for a measure of $\beta$, assuming it is not very small. Not an easy task due to the 
strong suppression of the small ratio of left to right charged gauge boson masses. 

Likewise, the couplings of the heavy doublet to quarks are readily seen to depend explicitly on $a$ and $\beta$, allowing for their simultaneous determination. 
They are also a function of the quark mixing matrix $V_R^q$ whose analytic dependance on $s_a t_{2\beta}$ can be found in~\cite{Senjanovic:2014pva}. The measurement of the RH quark mixing itself from $W_R$ interactions can thus also be used to deduce the essential CP-violating parameter $s_a t_{2\beta}$.

\Sec{VI. Conclusions and Outlook.}
In the SM the knowledge of charged fermion masses uniquely predicts the branching ratios of the Higgs boson decays. As shown in~\cite{Nemevsek:2012iq}, exactly the same happens in the minimal Left-Right Symmetric Model for the masses of light and heavy neutrinos with generalized charge conjugation as the LR symmetry. 
The case of generalized parity is more complex. We  obtained an explicit analogous result for purely hermitian Dirac Yukawa couplings and at the same time we have set the mathematical framework needed to address the general case.

  Moreover, in this work we devised a novel strategy to deal with this complex issue by exploiting the di-lepton final states in the decay of the LH doubly charged scalar. We showed that the flavor structure of these decays depends on the neutrino Dirac Yukawa couplings, which  can be determined from the decays of RH neutrinos into LH charged leptons.   There are additional processes  such as the leptonic decays 
of the heavy scalar doublet which probe these couplings as well. In particular, the decay of its neutral component into charged leptons pairs does this in a  transparent manner. 

Furthermore, we have demonstrated that only the hermitian part of the Dirac Yukawa couplings matters. This feature substantially facilitates its experimental determination by effectively halving the degrees of freedom to be measured. 

The complete determination of the neutrino Dirac mass matrix is admittedly a task for a future collider, but then so is the verification of the Higgs origin of light charged fermion masses.  Nonetheless, the point we are making is also one of principle, i.e., the neutrino Dirac Yukawa couplings in this theory correlate seemingly disconnected physical processes. This is what sets this theory apart from the usual seesaw mechanism in the SM.

We cannot overemphasize the fact that the LR symmetry used originally to understand parity violation in weak interactions
not only requires the existence of RH neutrinos which leads to the seesaw mechanism, but also connects the seesaw with a number of different physical processes and makes the LRSM a self-contained theory of neutrino mass.

\Sec{Acknowledgments.} We thank A. Melfo for useful discussions, comments and careful reading of the manuscript. GS acknowledges warm hospitality of the Stermasi establishment on the island of Mljet during the last stages of this work. 


\end{document}